\documentclass[12pt]{article}

\usepackage{graphicx}
\usepackage{epsf}

\usepackage{amsmath,amssymb,graphicx, float} 

\usepackage{pstricks}
\usepackage{color}
\usepackage{cite}

\def\beq{\begin{equation}}
\def\eq{\end{equation}}
\def\eeq{\end{equation}}

\def\centeron#1#2{{\setbox0=\hbox{#1}\setbox1=\hbox{#2}\ifdim
\wd1>\wd0\kern.5\wd1\kern-.5\wd0\fi
\copy0\kern-.5\wd0\kern-.5\wd1\copy1\ifdim\wd0>\wd1
\kern.5\wd0\kern-.5\wd1\fi}}
\def\ltap{\;\centeron{\raise.35ex\hbox{$<$}}{\lower.65ex\hbox{$\sim$}}\;}
\def\gtap{\;\centeron{\raise.35ex\hbox{$>$}}{\lower.65ex\hbox{$\sim$}}\;}

\def\chii0{\chi_i^0}
\def\chij0{\chi_j^0}

\def\foursqr#1#2{{\vcenter{\vbox{
 \hrule height.#2pt
 \hbox{\vrule width.#2pt height#1pt \kern#1pt
 \vrule width.#2pt}
 \hrule height.#2pt
 \hrule height.#2pt
 \hbox{\vrule width.#2pt height#1pt \kern#1pt
 \vrule width.#2pt}
 \hrule height.#2pt
     \hrule height.#2pt
 \hbox{\vrule width.#2pt height#1pt \kern#1pt
 \vrule width.#2pt}
 \hrule height.#2pt
     \hrule height.#2pt
 \hbox{\vrule width.#2pt height#1pt \kern#1pt
 \vrule width.#2pt}
 \hrule height.#2pt}}}}
\def\psqr#1#2{{\vcenter{\vbox{\hrule height.#2pt
 \hbox{\vrule width.#2pt height#1pt \kern#1pt
 \vrule width.#2pt}
 \hrule height.#2pt \hrule height.#2pt
 \hbox{\vrule width.#2pt height#1pt \kern#1pt
 \vrule width.#2pt}
 \hrule height.#2pt}}}}
\def\sqr#1#2{{\vcenter{\vbox{\hrule height.#2pt
 \hbox{\vrule width.#2pt height#1pt \kern#1pt
 \vrule width.#2pt}
 \hrule height.#2pt}}}}

\def\figin{\epsfcheck\figin}\def\figins{\epsfcheck\figins}
\def\epsfcheck{\ifx\epsfbox\UnDeFiNeD
\message{(NO epsf.tex, FIGURES WILL BE IGNORED)}
\gdef\figin##1{\vskip2in}\gdef\figins##1{\hskip.5in}
\else\message{(FIGURES WILL BE INCLUDED)}%
\gdef\figin##1{##1}\gdef\figins##1{##1}\fi}
\def\DefWarn#1{}
\def\figinsert{\goodbreak\midinsert}
\def\ifig#1#2#3{\DefWarn#1\xdef#1{fig.~\the\figno}
\writedef{#1\leftbracket fig.\noexpand~\the\figno}%
\figinsert\figin{\centerline{#3}}\medskip\centerline{\vbox{\baselineskip12pt
\advance\hsize by -1truein\noindent\footnotefont{\bf
Fig.~\the\figno:\ } \it#2}}
\bigskip\endinsert\global\advance\figno by1}

\def\fig#1#2#3#4{\vskip 0.5cm \begingroup \midinsert \centerline{
\psfig{file=#1,width=#2}} \vskip 0.4cm
\global\advance\figno by 1
\centerline{\vbox{\baselineskip=12pt \noindent Figure \the\figno: #3}}
\endinsert \endgroup {\xdef#4{\the\figno}} }

\def\figcrop#1#2#3#4#5#6#7#8{\vskip 0.5cm \begingroup \midinsert \centerline{
\psfig{file=#1,width=#2,bbllx=#3,bblly=#4,bburx=#5,bbury=#6}} \vskip 0.4cm
\global\advance\figno by 1
\centerline{\vbox{\baselineskip=12pt \noindent Figure \the\figno: #7}}
\endinsert \endgroup {\xdef#8{\the\figno}} \vskip .5cm}

\def\figlabel#1{\xdef#1{\the\figno}}
\def\encadremath#1{\vbox{\hrule\hbox{\vrule\kern8pt\vbox{\kern8pt
\hbox{$\displaystyle #1$}\kern8pt}
\kern8pt\vrule}\hrule}}
\def\underarrow#1{\vbox{\ialign{##\crcr$\hfil\displaystyle
 {#1}\hfil$\crcr\noalign{\kern1pt\nointerlineskip}$\longrightarrow$\crcr}}}

\usepackage{mathrsfs}
\usepackage{mathbbol}
\def \bea{\begin{eqnarray}}
\def \eea{\end{eqnarray}}
\def \ba{\begin{eqnarray*}}
\def \ea{\end{eqnarray*}}
\def\nn{\nonumber}

\begin{document}

\begin{titlepage}

\begin{center}
\vspace*{-1cm}

\hfill RU-NHETC-2014-21\\
\vskip 1.0in
{\LARGE \bf Higgs Boson Yukawa Form Factors from} \\

\vspace{.25in}
{\LARGE \bf 
Supersymmetric Radiative Fermion Masses }\\

\vspace{.15in}

\vskip 0.45in
~~{\large Arun Thalapillil} ~~and~~
{\large Scott Thomas}

\vskip 0.25in

{\em New High Energy Theory Center} \\
{\em  Department of Physics} \\
{\em Rutgers University}   \\
{\em Piscataway, NJ 08854 }

\vspace*{1in}

\vskip 0.145in

\end{center}

\baselineskip=16pt

\begin{abstract}
\noindent 
The recent discovery of the Higgs-like resonance at $125\,\rm{GeV}$ has opened up new avenues in the search for beyond standard model physics. Hints of such extensions could manifest themselves as modifications in the Higgs-fermion couplings and other Higgs related observables. In this work, we study aspects of a class of models where the light fermion masses are radiatively generated. Specifically, we consider models where the light fermion masses, partially or completely, arise from chiral violation in the soft supersymmetry-breaking sector. In these models, the radiatively generated Higgs-fermion Yukawa form factors have non-trivial characteristics and will modify Higgs-fermion couplings from their standard model expectations. A radiatively generated fermion mass could also potentially contribute to large anomalous magnetic moments; this is particularly interesting in the case of the muon where a persistent discrepancy, at the level of around $3\,\sigma$, has existed between experiment and theory. Deviations in the Higgs-fermion couplings will eventually be probed to high accuracy in the near future, at the LHC and the planned ILC, to less than a percent. The prospect of a large, unknown contribution to the muon anomalous magnetic moment could be reaffirmed as well, in future experiments.  All these reasons make it worthwhile to revisit models of radiatively generated fermion masses, and investigate some of their general characteristics in these contexts.
\end{abstract}

\end{titlepage}

\baselineskip=17pt

\newpage






\section{Introduction} 
\par
The origin of the distinctive hierarchical pattern of quark and lepton Yukawa couplings to the Higgs condensate has remained a mystery since the elucidation of the Standard Model of particle physics many decades ago. With the recent discovery of a Higgs-like boson at around $125\,\rm{GeV}$, by both the ATLAS and CMS collaborations~\cite{{Chatrchyan:2012ufa},{Aad:2012tfa}}, we have entered a new era for particle physics. Probing the couplings of this new Higgs-like resonance could reveal deviations from standard model (SM) expectations and could present new opportunities for the discovery of beyond standard model physics (BSM). This is especially pertinent given the lack of evidence, so far, for BSM physics in other search channels. An intriguing possibility is that the fermion masses, partially or completely, may be radiatively generated~\cite{Weinberg:1972ws}. There have been numerous realizations of such models in the literature~\cite{Barr:1979xt,Barbieri:1980tz,Ibanez:1981nw,Barbieri:1981yy,Lahanas:1982et,Nanopoulos:1982zm,Masiero:1983ph, delAguila:1984qs, Banks:1987iu, Kagan:1987tpa,Balakrishna:1987qd,Balakrishna:1988ks,Balakrishna:1988bn,Ma:1988qc,Ma:1989tz,He:1989er,Kagan:1989fp,ArkaniHamed:1995fq,ArkaniHamed:1996zw,Hamzaoui:1998yy,Babu:1998tm,Borzumati:1997ne,Borzumati:1997bd,Borzumati:1999sp,DiazCruz:2000mn,Ferrandis:2004ng,Ferrandis:2004ri,Crivellin:2008mq,Crivellin:2010ty,Crivellin:2011sj,ArkaniHamed:2012gw,Baumgart:2014jya,Berezhiani:1991ds,Dobrescu:2008sz,Hashimoto:2009xi,Graham:2009gr,Ibarra:2014fla,Joaquim:2014gba,Altmannshofer:2014qha, Fraser:2014ija}. 

\par
In supersymmetric theories, the chiral symmetry responsible for fermion masses may be broken by hard renormalizable terms or by soft supersymmetry breaking terms. The latter is a possibility since the SM fermions carry the same flavor quantum-numbers as the sparticles. In this work, we will focus on supersymmetric models where the chiral flavor symmetry is broken, partially or completely, by auxiliary expectation values arising from soft, dimension-three, supersymmetry breaking terms in the lagrangian~\cite{Borzumati:1997ne,Borzumati:1997bd,Borzumati:1999sp}. More specifically- The fermion chiral and $\text{U(1)}_R$ symmetries are broken by the gaugino masses, while the chiral flavor symmetries are broken by the tri-linear A-terms. Note that since the fermion and scalar R-charges differ by unity, as $R_\psi=R_\phi-1$, the A-term and Yukawa background spurions have different R-charges. Thus, radiative fermion masses in supersymmetry require $\text{U(1)}_R$ breaking along with fermion chirality breaking. Following~\cite{Borzumati:1997ne,Borzumati:1997bd,Borzumati:1999sp}, we will call these models soft-Yukawa models.
\par
Such radiatively generated fermion masses have marked characteristics related to their soft origin. We investigate two of these features in particular. One of the consequences is that the fermion couplings to the Higgs boson is modified by non-trivial form factors and will appear, in Higgs branching-fractions for instance, as deviations from the usual SM expectations. This has been an active line of enquiry in general supersymmetric theories~\cite{Higgsferfer}. We wish to revisit this in our context. Another characteristic we investigate is that in these models the fermion anomalous magnetic moments could be large. This is a consequence of the fact that both the fermion mass and the anomalous magnetic moment arise at the same loop-order. Due to this, one does not pay any extra price in terms of loop-factors and the contribution is therefore effectively a loop-factor larger than with tree-level Yukawas.
\par
The emphasis in this study will be to glean general features of the soft-Yukawa models pertaining to their contributions to Higgs-fermion Yukawa form factors and fermion anomalous magnetic moments. In this context we will, as far as possible, work in a simplified framework where flavor-mixing effects and CP-violation will be ignored. With a non-trivial flavor structure or new CP-phases, it is possible that consequences could also show up in flavor observables in the near future. In a recent study~\cite{Altmannshofer:2014qha}, for instance, it was attempted to construct the full flavor hierarchy of the standard model, in such a soft-Yukawa framework. We will update and generalize some aspects of the studies in~\cite{Borzumati:1997ne,Borzumati:1997bd,Borzumati:1999sp}.
\par
In the next section we set the stage for our study, by first defining the relevant dimension-six operators through which we will parametrize our deviations from the SM expectations. We then proceed in section \ref{sec:dimsixcouplingscomp} to derive explicitly the expression for the radiatively generated fermion mass and coefficients of the dimension-six operators, in the soft-Yukawa model. Then, in section \ref{sec:YukawaFormFactor} we will put together all these ingredients to study deviations in the Higgs-fermion couplings, arising from these operators. Subsequently, in section \ref{sec:AnomalousMagneticMoment} we will consider the dimension-six fermion electromagnetic dipole operator and investigate contributions to the fermion anomalous magnetic moments, specifically the muon. Finally, in section \ref{sec:conclusions} we summarize our results and conclude.


\section{Dimension-Six Yukawa Form Factor Operators } 
\label{sec:dimsixcouplings}

\par
The potential deviations of the Higgs couplings to fermions may be parametrized through non-renormalizable, dimension six operators that are invariant under the SM gauge group~\cite{{Buchmuller:1985jz},{Grzadkowski:2010es}}. We will use this framework to investigate the deviation of the Higgs-fermion couplings in the soft-Yukawa model.
\par
 The main terms of interest to us in the Lagrangian, in our context, are the dimension-four linear Yukawa coupling operator
 
 \beq
- \lambda_f ~\bar{f} H  f ~+~ {\rm h.c.}
\label{dim4lyukawa}
\eq
and the dimension-six cubic Yukawa and Yukawa-radius operators\\
\par
\beq
- \eta_f ~H^\dagger H ~ \bar{f} H f ~-~ \zeta_f ~\bar{f} D^2 H  f ~+~ {\rm h.c.}
\label{dim6ffhops}
\eq
Here $D$ is a covariant derivative and $f,\,\bar{f}$ are two-component fermions in the Weyl basis.
\par
In a general analysis with dimension-six operators, there is also a Lagrangian operator
\beq
 \partial_\mu \lvert H \rvert^2 \partial^\mu \lvert H \rvert^2
\eq
that modifies Higgs-fermion couplings, through wavefunction renormalizations of the physical Higgs field~\cite{Giudice:2007fh,Craig:2013xia}. This multiplicative renormalization will effectively be a higher order correction in the soft-Yukawa framework we consider and is therefore not included here.
\par
In principle, the Yukawa radius operator $ \zeta_f ~\bar{f} D^2 H  f $ may be eliminated, using the equations of motion, in favor of the linear and cubic Yukawa operators with coefficients~\cite{CraigThomasThalapillil}
\beq
\lambda_f' = \lambda_f + {m_h^2 \over 2} ~\zeta_f 
\eq
\beq
\eta_f' = \eta_f - {m_h^2 \over v^2} ~\zeta_f 
\eq
But note that the combination $(v^2 \eta_f - m_h^2 \zeta_f)$ is independent 
of the operator basis, and immune to the choice of whether the Yukawa-radius term is retained or eliminated using the equations of motion. We shall see shortly that it is indeed this basis independent combination of terms that appear
in the ratio of the Higgs to mass Yukawa couplings (which we will call $\kappa$ below). This ratio will parametrize the deviation of the Higgs-fermion couplings from SM values uniquely, and may therefore be considered as an effective Higgs-Yukawa form factor. 
\par
Now, the net fermion mass operator with broken 
electroweak symmetry defines an effective fermion mass Yukawa coupling
\beq
 - m_f \bar{f} f ~+~ {\rm h.c.}
 \equiv - { \lambda_f^{\text{\tiny{ eff}}} \over \sqrt{2} } ~v \bar{f} f ~+~ {\rm h.c.} 
\eq
Similarly, the operator giving the net coupling of an on-shell Higgs boson to fermions defines an effective Higgs Yukawa coupling 
\beq
 -{\lambda^h_f \over \sqrt{2} } ~h \bar{f} f ~+~ {\rm h.c.} 
\eq
In our conventions, the Higgs field with canonically normalized expectation value 
and the Higgs boson are all related by
\beq
H = {1 \over \sqrt{2} } \left( v + h \right)  
\eq
Note that the mass Yukawa coupling may equivalently be defined
formally through the low energy Higgs theorem 
\beq
{ \lambda_f^{\text{\tiny{ eff}}} \over \sqrt{2} } = {\partial m_f \over \partial v} 
\eq
\par
The important point to consider is that in the renormalizable theory, with only marginal interactions
at tree-level, the fermion-mass and Higgs-boson Yukawa couplings are equal $ \lambda_f^{\text{\tiny{ eff}}} = \lambda^h_f \equiv \lambda_f$ with $\lambda_f$ as defined is Eq.\,(\ref{dim4lyukawa}). However, in the presence of the dimension-six operators of Eq.\,(\ref{dim6ffhops}), the equality between these two effective 
Yukawa couplings is lifted and in general we have $ \lambda_f^{\text{\tiny{ eff}}} \neq \lambda^h_f $.
\par
With the terms in Eq.\,(\ref{dim6ffhops}), the mass-Yukawa receives contributions from both the linear and cubic Yukawa 
couplings, while the Higgs-Yukawa receives contributions from the linear Yukawa, cubic Yukawa, as well as the Yukawa radius couplings
\begin{eqnarray} 
 \lambda_f^{\text{\tiny{ eff}}} &=& \lambda_f + {v^2 \over 2}  \eta_f   \nn \\ 
\lambda^h_f &=& \lambda_f + {3 v^2 \over 2}  \eta_f  - m_h^2  \zeta_f 
 \label{effmyukeffhyuk}
\end{eqnarray}
Note the relative factor of $3$ prefacing the cubic Yukawa term in the Higgs-Yukawa coupling, coming from the different combinatorics. The $m_h^2$ prefacing the Yukawa radius coupling is due to the equation of motion; thus the Higgs field with this coupling is assumed on-shell.
\par
The ratio of the Higgs to mass Yukawa couplings is then given by
\beq
\kappa_f \equiv { \lambda^h_f \over  \lambda_f^{\text{\tiny{ eff}}}} = 
 1 + {v^2 \eta_f  - m_h^2 \zeta_f  \over  \lambda_f^{\text{\tiny{ eff}}} }  
 \label{kappaops}
\eq
The deviation of this ratio from unity, induced by the dimension-six cubic Yukawa and Yukawa radius operators, represent a net Yukawa form factor for the Higgs-fermion interactions. This quantity will parametrize the deviation of the Higgs-fermion couplings from their SM values. As we alluded to before, note that it is the combination ${v^2 \eta_f  - m_h^2 \zeta_f }$ that appears here.
\par
In the next section we will define some of our conventions and then proceed to compute the topologies which we require to study the Higgs-fermion form factor and contributions to the anomalous magnetic moment.

\section {Radiative Fermion Masses \& Yukawa Form Factors in Supersymmetric Theories}

\label{sec:dimsixcouplingscomp} 

\begin{figure}[htb]
\begin{center}
\includegraphics[width=8.0cm]{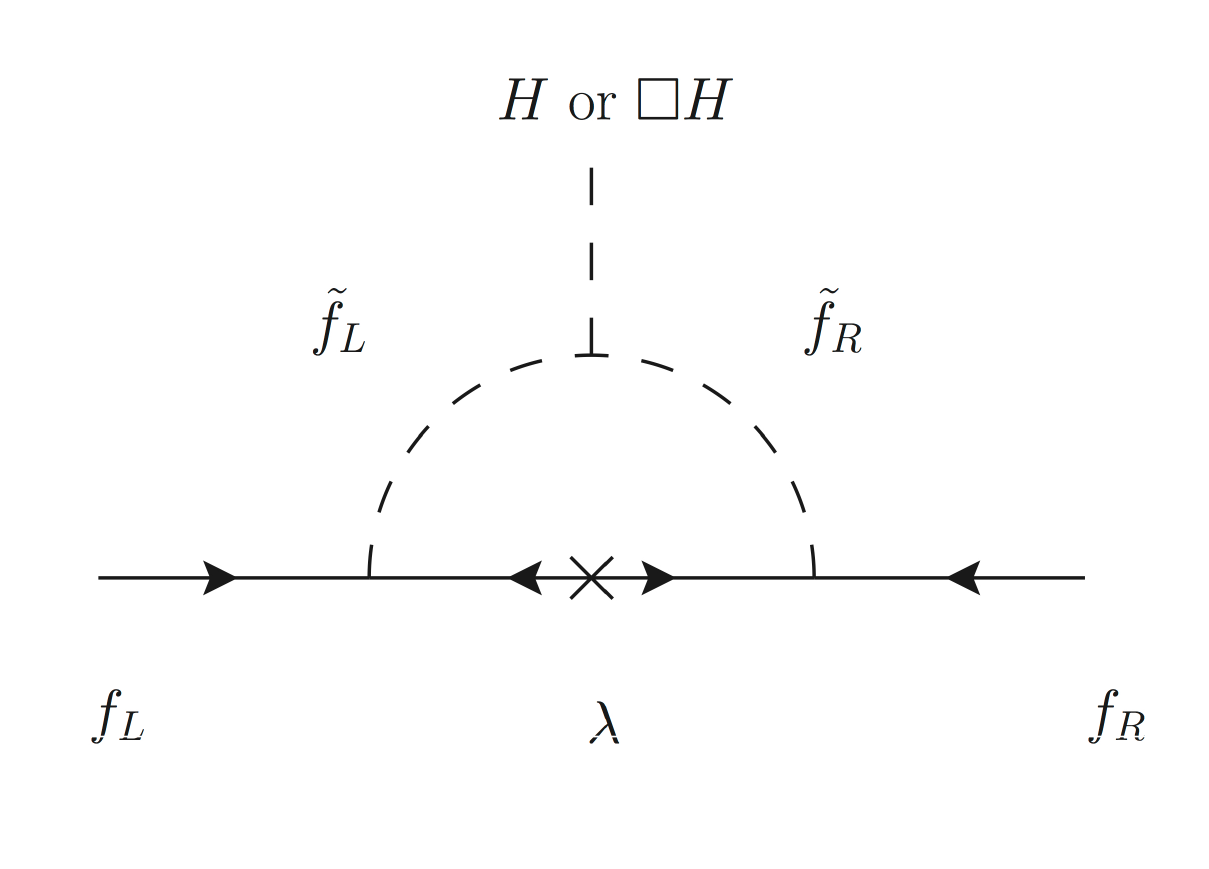} 
\caption{One-loop supersymmetric contributions to the chirality violating Yukawa operator. Arrows indicate fermion chirality, and the Hermitian conjugate operators are not shown. Chiral and flavor violation comes from the gaugino mass and tri-linear $A$-term respectively. The gaugino mass also provides the necessary $U(1)_R$ breaking.}
\label{fig:LinearYukop}
\end{center}
\end{figure}

\begin{figure}[htb]
\includegraphics[width=8.25cm]{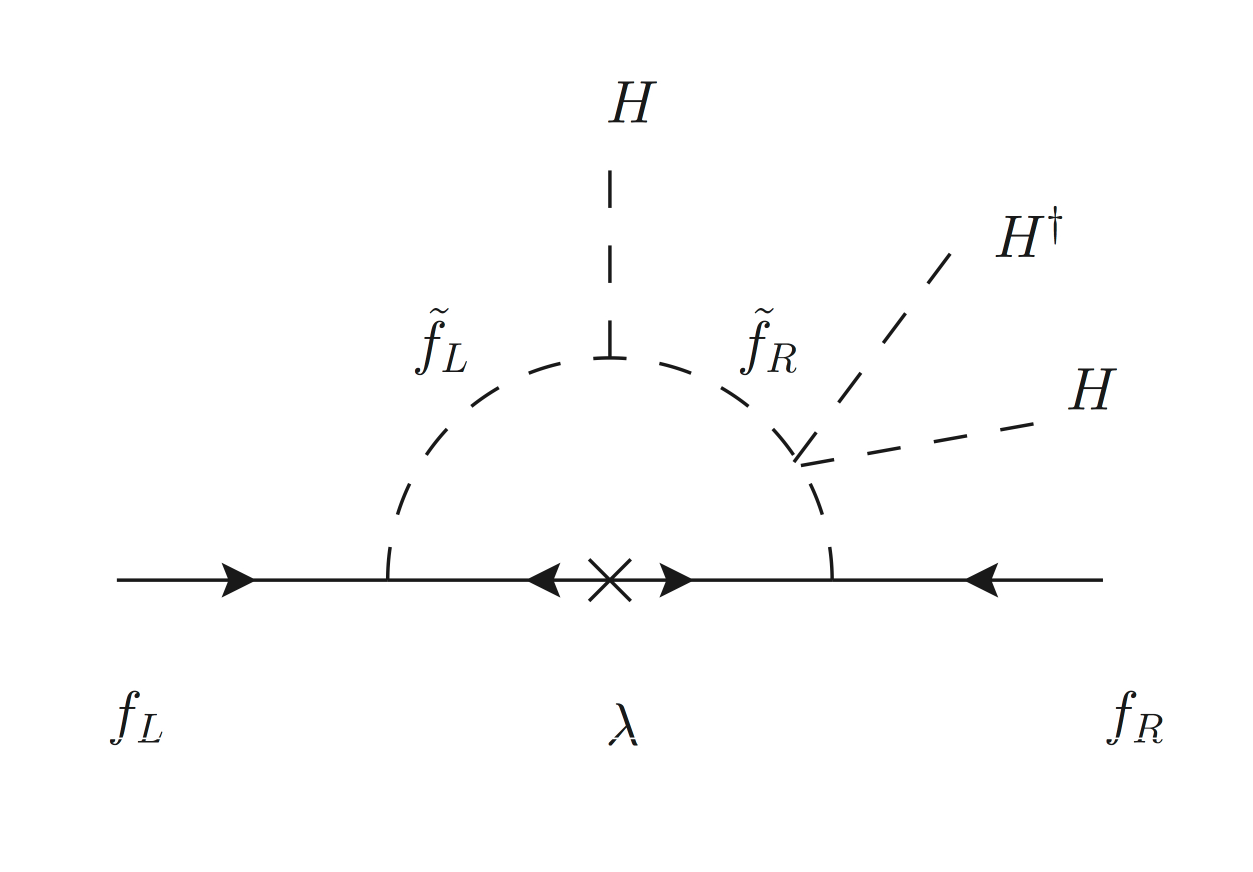} 
\includegraphics[width=7.75cm]{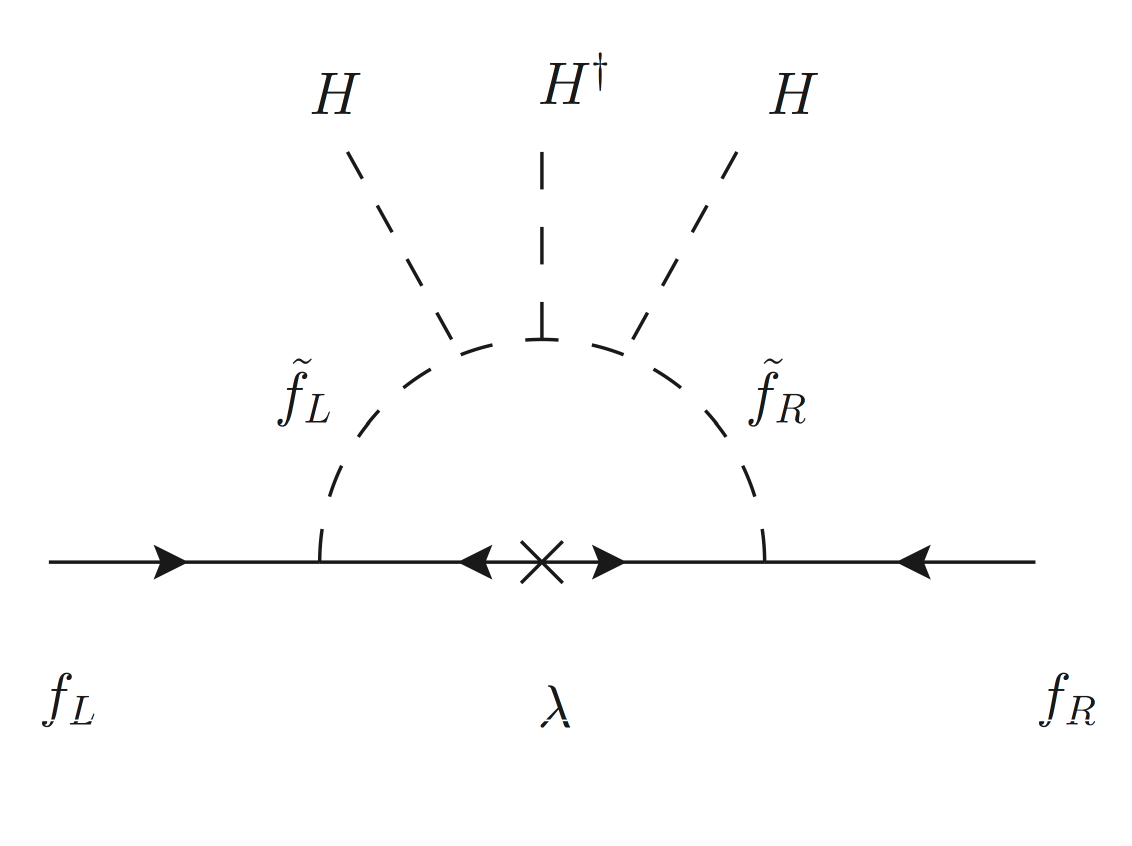} 
\caption{One-loop supersymmetric contributions to the chirality violating cubic Yukawa operator. Non-linear Higgs interactions come from $D$-term scalar gauge interactions and terms proprtional to $A_f^3$. Diagrams with $D$-term interactions between Higgs and left-handed fermion superpartners are not shown. Again, arrows indicate fermion chirality.}
\label{fig:CubicYukop}
\end{figure}


The specific UV-completion that we focus on that could potentially lead to sizable deviations in the Higgs-fermion couplings, among other effects, is one where the fermion masses are generated through radiative processes~\cite{Weinberg:1972ws, Barr:1979xt,Barbieri:1980tz,Ibanez:1981nw,Barbieri:1981yy,Lahanas:1982et,Nanopoulos:1982zm,Masiero:1983ph, delAguila:1984qs, Banks:1987iu, Kagan:1987tpa,Balakrishna:1987qd,Balakrishna:1988ks,Balakrishna:1988bn,Ma:1988qc,Ma:1989tz,He:1989er,Kagan:1989fp,ArkaniHamed:1995fq,ArkaniHamed:1996zw,Hamzaoui:1998yy,Babu:1998tm,Borzumati:1997ne,Borzumati:1997bd,Borzumati:1999sp,DiazCruz:2000mn,Ferrandis:2004ng,Ferrandis:2004ri,Crivellin:2008mq,Crivellin:2010ty,Crivellin:2011sj,ArkaniHamed:2012gw,Baumgart:2014jya,Berezhiani:1991ds,Dobrescu:2008sz,Hashimoto:2009xi,Graham:2009gr,Ibarra:2014fla,Joaquim:2014gba,Altmannshofer:2014qha,Fraser:2014ija} in a supersymmetric framework.
\par
Specifically, from a physically viable point of view, we will consider a simple supersymmetric model where such a mechanism is responsible for fermion masses in the first, second generation fermions and probably also the bottom-quark. For instance, consider the topologies of Figs.\,\ref{fig:LinearYukop} and \ref{fig:CubicYukop}. They could lead to the dimension-six operators of Eq.\,(\ref{dim6ffhops}), when supersymmetry is softly broken. The topology with a single Higgs insertion, of Fig.\,\ref{fig:LinearYukop}, gives a contribution to the linear Yukawa, $\lambda_f$, and a contribution to the Yukawa-radius, $\zeta_f$, for non-zero external Higgs momenta. In addition, with an additional photon insertion this topology also leads to the fermion electromagnetic dipole operator, $\xi_f$, that we shall discuss in sec.\,\ref{sec:AnomalousMagneticMoment}. The topology with  three Higgs insertions, of Fig.\,\ref{fig:CubicYukop}, leads to the cubic Yukawa, $\eta_f$. The contribution to the cubic Yukawa are due to D-term interactions and terms proportional to $A_f^3$, where $A_f$ is the soft tri-linear term.
\par
Let us define some of our other conventions before proceeding. Our Lagrangian soft terms are defined as
\beq
  -A_u ~ \tilde{Q} H_u \tilde{\bar{u}} 
- A_d ~ \tilde{Q} H_d \tilde{\bar{d}} 
- A_\ell ~ \tilde{L} H_d \tilde{\bar{e}} ~+~{\rm h.c.} 
\eq
We only consider holomorphic trilinear couplings for simplicity. In the alignment limit, we can approximate $H_u=H\sin\beta$ and $H_d=H\cos\beta$. Here, $H$ is the standard model Higgs field and $\tan\beta=\langle H_u \rangle /\langle H_d \rangle$. In this limit, the above may then be re-written as
\beq
- \tilde{A}_u ~ \tilde{Q} H \tilde{\bar{u}} 
- \tilde{A}_d ~ \tilde{Q} H \tilde{\bar{d}} 
- \tilde{A}_\ell ~ \tilde{L} H \tilde{\bar{e}} ~+~{\rm h.c.} 
\eq
with the re-definitions $ \tilde{A}_u =  ~A_u\, \sin\beta \,,\,\tilde{A}_d =~A_d\,  \cos \beta$ and $\tilde{A}_\ell =~A_\ell \,  \cos \beta $. 
\par
In our conventions, the LR-mixing term in the squark/slepton mass-squared matrix has the form
\beq
m_{LR}^2 =  \tilde{A}_f \langle H^0 \rangle \equiv \tilde{X}_f  
\label{LRterm}
\eq
where there is no term proportional to the Higgsino mass $\mu$, since the tree-level Yukawa couplings are assumed to be absent or very small ab initio.
\par
We will work in a simplified, $2\times 2$ limit for the sparticle mass-squared matrices, to obtain simple analytic expressions. The generalization of the expressions to a full three-family case should in principle be straightforward, but with more complicated analytic expressions and mixing matrix factors. In the simple limit we are considering, the corresponding mixing angle is simply given by $\sin 2 \theta_f = - {2 \tilde{X}_f / (m_2^2 - m_1^2) } $. The LR-mixing terms will appear in our analytic expressions due to algebraic identities between mixing angle factors. The mass eigenstates, for our case, in this approximation are given by 
\beq
m_{1,\, 2}^2=\frac{1}{2}\left[m_L^2+m_R^2+\Delta_L+\Delta_R \pm \sqrt{(m_L^2-m_R^2+\Delta_L-\Delta_R)^2+ 4 \tilde{X}_f^2 }\right] 
\eq
Here, $m_L$ and $m_R$ are the left and right sparticle mass soft-terms in the $2\times 2$ mass-squared matrix; $\Delta_L$ and $\Delta_R$ are the D-term contributions given by $\Delta_L=(T^f_{3L}-Q^f s_w^2)\,m_Z^2 \cos 2\beta$ and $\Delta_R=Q^f s_w^2\,m_Z^2 \cos 2\beta$. $s_w^2$ is the weak-angle, $m_Z$ is the Z-boson mass, $T^f_{3L}$ is the third component of the weak-isospin from $SU(2)_L$ and $Q_f$ is the electromagnetic charge.

\par
We will also be working to leading order in $m^2_{\text{\tiny{LR}}}/\tilde{m}^2_{\text{\tiny{SUSY}}}$, to obtain expressions for the dimension-six coefficients in Eq.\,(\ref{dim6ffhops}). This approximation should be justified so long as $\tilde{A}v/\tilde{m}_{\text{\tiny{SUSY}}}^2\ll 1$ and will capture almost all the salient features that we are interested in. Finally, we will set the left and right elements of the $2\times2$ squark/slepton mass-squared matrix at the end to be equal, $m_L=m_R$, to make plots.
\par
Let us now proceed to calculate each of the coefficients in Eq.\,(\ref{dim6ffhops}), from the topologies of Fig.\,\ref{fig:LinearYukop} and Fig.\,\ref{fig:CubicYukop}. We will then finally put all the components together, to investigate the deviation of Higgs-fermion couplings and the contribution to the anomalous magnetic moment in sections \ref{sec:YukawaFormFactor} and \ref{sec:AnomalousMagneticMoment}.


 \subsection{Radiative Fermion Masses}
 \par
 Let us begin by computing the radiatively generated mass term. The radiatively generated mass and linear, mass-Yukawa coupling arise from the topology in Fig.\,\ref{fig:LinearYukop}, where the Higgs field is replaced with a vacuum expectation value (VEV). The linear Yukawa coupling arises from the dimensionful soft parameters and are finite. They are therefore soft in the technical sense, requiring no counter terms.
 \par
 From the topology of Fig.\,\ref{fig:LinearYukop}, the two-point, chirality-violating function at zero external momenta, gives the radiative fermion mass
\beq
m_f=\bigg[ ~ 
i \int d^4x~ \langle 0 | T \{ f(x)  \bar{f}(0) \} | 0 \rangle 
~ \bigg]^{-1}
=  {1 \over 16 \pi^2} ~ g^2 C ~ { \tilde{X}_f   \over m_\lambda } ~ 
f_\lambda(m_1^2 / m_\lambda^2 , m_2^2 / m_\lambda^2) 
\label{massCoupling}
\eq
Here, $\tilde{X}_f$ is as defined in Eq.\,(\ref{LRterm}), $C$ is an appropriate color or hypercharge factor depending on the gaugino under consideration. For instance, $C=4/3$ for the gluino and $C=-1/2$ for the bino. Note that we have used algebraic identities to represent the mixing angle coefficients in terms of $m^2_{\text{\tiny{LR}}}$. The loop function appearing above is defined as 
\beq
f_\lambda(x,y)  =  {g(x) - g(y)  \over  x-y }  
\eq
where $g(x)= 2 x \ln x /(x-1)$. In the degenerate limit
\beq
f_\lambda(x,x) = { 2( x - 1 - \ln x) \over (1-x)^2 }  
\eq
Also note that $f_\lambda(x,y) = f_\lambda(y,x)$ and with our normalization $f_\lambda(1,1)=1$.
\par
We may assume that the radiative quark masses are generated dominantly by gluino exchange, which is enhanced by $\alpha_s$ and in which case there is a sizable color factor, $C=4/3$. In the case of leptons we may assume without loss of generality that it is the lightest-bino, $\tilde{B}$, that is giving the dominant contribution. Note that $\tilde{W}_3-\tilde{W}_3$ exchange in the loop is disallowed, since the loop contains both left and right-handed sparticles. $\tilde{B}-\tilde{W}_3$ is parametrically suppressed by mixing matrix factors in both the pure gaugino or pure higgsino limits. Pure Higgsino and chargino contributions are also disallowed in the absence of tree-level Yukawas. Finally, note also that in the absence of hard tree-level Yukawa couplings, there are no contributions from LR-mixing terms proportional to the Higgsino mass term $\mu$.
\par
It is intuitively clear that the first-generation fermion masses, being tiny, may be typically accommodated for small values of the tri-linear term. For intermediate to large A-terms one could speculate the possibility of the second-generation fermion masses to also be included. For the third-generation, not surprisingly, it is very unlikely that the dominant component of their masses are being generated radiatively while keeping the vacuum stability conditions intact- which we discuss shortly. Nevertheless, we will see that a radiatively generated b-quark mass is still possible.

\begin{figure}[p!]
\begin{center}
\includegraphics[width=12cm]{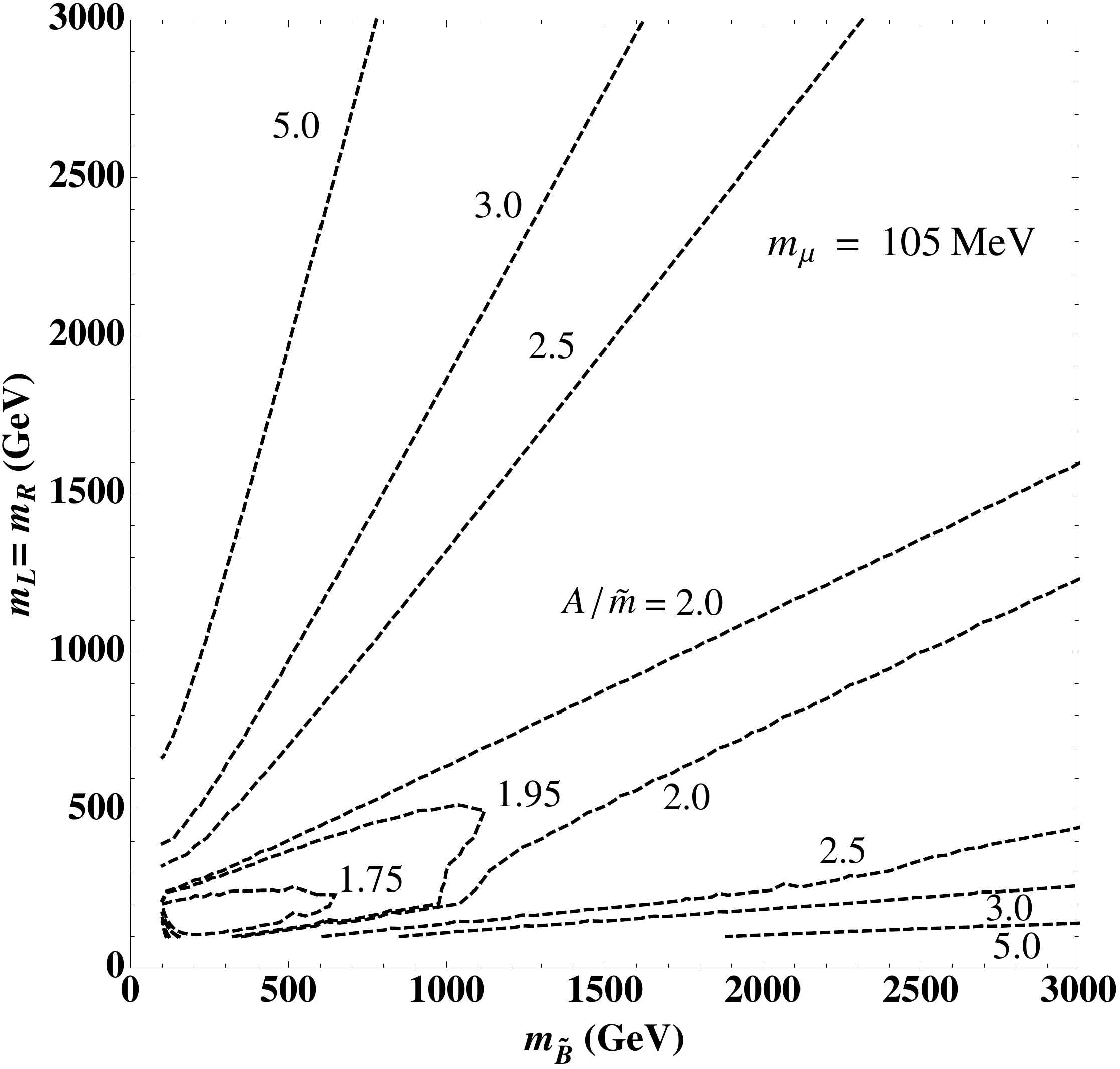} 
\caption{ Contours of fixed $A/\tilde{m}$ illustrating viable regions for the radiative generation of the muon mass. The plot is shown for $\tan\beta=1$. Here, $\tilde{m}^2=\frac{1}{2}\left(m_1^2+m_2^2\right)={1 \over 2 } (m_L^2+m_R^2)$ and the muon mass at the electroweak scale was taken to be $105\,\rm{MeV}$. Note that values of up to $A/\tilde{m} \sim \mathcal{O}(2)$, for a Higgs mass around $125\,\rm{GeV}$, satisfy the metastability bound $A/\mathfrak{m}~\lesssim~1.75$.  For larger values of $\tan\beta$, much greater than unity, due to suppression by the $\cos\beta$ factor, it becomes more difficult to get a viable muon mass satisfying the metastability limit.
}
\label{fig:AbyMMuon}
\end{center}
\end{figure}

\begin{figure}[p!]
\begin{center}
\includegraphics[width=12cm]{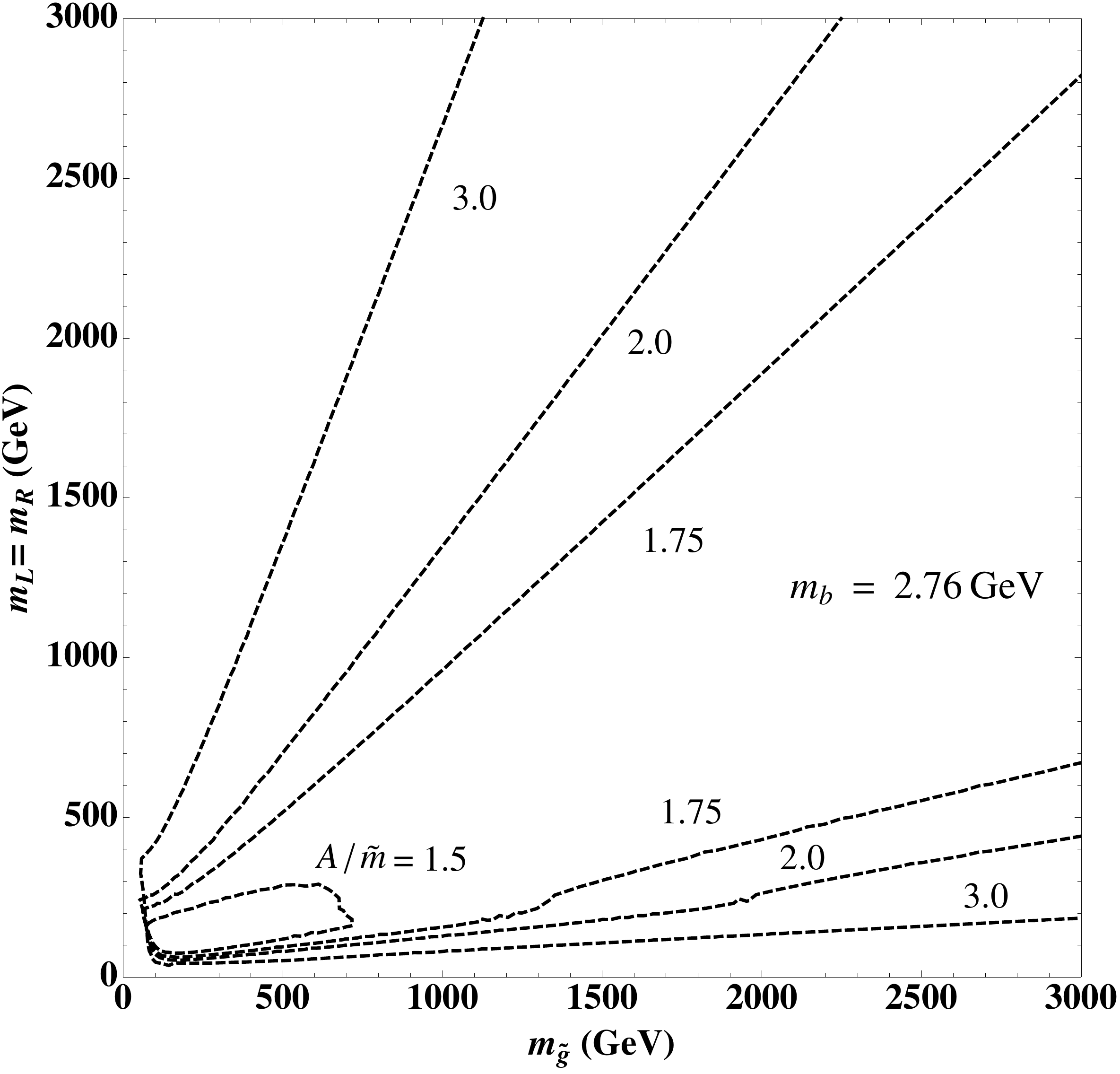} 
\caption{ Contours of fixed $A/\tilde{m}$ representing feasible regions for the radiative generation of $m_b$, again for $\tan\beta=1$. We have, as before, defined $\tilde{m}^2=\frac{1}{2}\left(m_1^2+m_2^2\right)$. The value $m_b=\,2.76\,\rm{GeV}$ at the electro-weak scale, was assumed to make the plots. Due to the large $\alpha_s$ coupling and color factor, there is a larger swath of viable parameter space compared to the muon. For larger values of $\tan\beta$, it becomes more difficult to get a viable mass satisfying the metastability limit again. Values up to $A/\tilde{m} \sim \mathcal{O}(2)$ once more would satisfy the metastability bound $A/\mathfrak{m}~\lesssim~1.75$ here.
}
\label{fig:AbyMBottom}
\end{center}
\end{figure}

\par
Let us consider in more detail two of the interesting cases. From Eq.\,(\ref{massCoupling}), the expressions for a radiatively generated muon mass and  bottom-quark mass may be written as 
\beq
m_\mu = -
{1 \over 32 \pi^2} ~ g^{'2} ~ { \tilde{A}_\mu  \over m_{\tilde{B}} } ~ 
f_\lambda(m_1^2 / m_{\tilde{b}}^2 , m_2^2 / m_{\tilde{b}}^2)~ { v \over \sqrt{2} }  
\label{muonmassCoupling}
\eq

\beq
m_b = 
{1 \over 12 \pi^2} ~ g_s^{2} ~ { \tilde{A}_b  \over m_{\tilde{g}} } ~ 
f_\lambda(m_1^2 / m_{\tilde{g}}^2 , m_2^2 / m_{\tilde{g}}^2)~ { v \over \sqrt{2} }  
\label{bottommassCoupling}
\eq

 In the muon case, the gaugino is assumed to be the lightest-bino, while in the bottom-quark case the gaugino is assumed to be the gluino. In Figs.\,\ref{fig:AbyMMuon} and \ref{fig:AbyMBottom}, we illustrate the regions for each of these cases, where one obtains a realistic value for the fermion mass. The figures are plotted with $\tan\beta=1$, for constant values of $A/\tilde{m}$ with $ \tilde{m}^2=\frac{1}{2}\left(m_1^2+m_2^2\right)=m_L^2$, after setting $m_L=m_R$. For a given value of $A/\tilde{m}$, the isoclines of Figs.\,\ref{fig:AbyMMuon} and \ref{fig:AbyMBottom} are approximately diagonal lines. This is understood from the fact that the defining locus of the constraint equations above are given by $k\,f_\lambda(k^2)=\text{const.}$, where $k= (\tilde{m}/m_\lambda) $. This may be approximated roughly to a quadratic expression with two solutions in general, for $k$. These give the two lines in the plot, for a fixed $A/\tilde{m}$.
\par
Let us address briefly the question of vacuum stability in soft-Yukawa models. In the soft-Yukawa models of radiatively generated fermion masses, one may require large trilinear A-terms to obtain realistic fermion masses, for instance in certain regions of Figs.\,\ref{fig:AbyMMuon} and \ref{fig:AbyMBottom}. This is potentially dangerous as large A-terms could lead to color or charge-breaking. Thus, it is important to investigate the allowed regions in the $(m_\lambda,\,\tilde{m})$ parameter space, which will provide an upper bound on the trilinear terms, that give viable fermion masses while also satisfying stability criteria. From detailed investigations~\cite{Borzumati:1999sp} it has been deduced that for absolute stability one requires $\frac{A}{\mathfrak{m}}~\ll~1$ and for metastability one requires $\frac{A}{\mathfrak{m}}~\lesssim~1.75$, where $\mathfrak{m}^2=\frac{1}{3}\left(m_L^2+m_R^2+m_H^2\right)$ and $m_H$ is the Higgs mass term. These bounds implicitly assume a minimal holomorphic scenario~\cite{Borzumati:1999sp} where all of the fermion mass is coming from the holomorphic Higgs coupling through the tri-linear A-term. In such models the quartic scalar coupling in the potential also comes out to be negative and small. The bounds are obtained from the condition that in the direction where the tri-linear term contribution to the potential is most negative, the global minima occurs at the origin. This is a sufficient condition to avoid charge and color breaking. We will use the above criteria to estimate constraints from vacuum stability on our simple model of soft-Yukawas. It is observed that the fermion masses, $m_\mu$ and $m_b$, may be viably obtained upto $A/\tilde{m} \sim \mathcal{O}(2)$ which for a Higgs mass around $125\,\rm{GeV}$ would comfortably satisfy $ A/\mathfrak{m}~\lesssim~1.75$. Note that in the case of $m_b$, due to the larger $\alpha_s$ coupling and color-factor, there is a wider swath of allowed parameter space for smaller values of $A/\tilde{m}$. For $\tan\beta \gg 1$, it becomes harder to get viable, radiatively generated muon and bottom-quark masses satisfying the metastability criterion. This is a consequence of the $\cos\beta$ suppression present in the constraint equations; remember that $\tilde{A}_{d/\ell}=A_{d/\ell}\cos\beta$

\subsection{ Linear Yukawa Coupling}
\par
Starting from the expression in Eq.\,(\ref{massCoupling}), for the radiatively generated fermion mass, we may now derive the expression for the linear Yukawa coupling defined by
$$
- \lambda_f ~\bar{f} H  f ~+~ {\rm h.c.}
$$
Using the Higgs low-energy theorem,
\beq
{ \lambda_f  \over \sqrt{2} } = {\partial m_f \over \partial v}\Big |_{v=0} 
\eq
we obtain the corresponding linear Yukawa coupling term, in Weyl basis, 
\beq
\lambda_f =  
{1 \over 16 \pi^2} ~ g^2 C ~ { \widetilde{A}_f \over m_\lambda } ~ 
f_\lambda(m_{L}^2 / m_\lambda^2 , m_{R}^2 / m_\lambda^2) 
\label{YukawaCoupling}
\eq
This may be thought of as the leading order term in an expansion, in terms of $\tilde{A}v/\tilde{m}_{\text{\tiny{SUSY}}}^2$. As previously mentioned, this is well defined for $v/\tilde{m}_{\text{\tiny{SUSY}}} \ll 1$. As we have seen,  $A/\tilde{m}_{\text{\tiny{SUSY}}}\sim \mathcal{O}(1)$, may be required to get viable muon and bottom-quark masses and the expansion is still valid in this limit.
\par
Note that the linear Yukawa coupling approaches a constant value, independent of the supersymmetry-breaking scale, as we decouple the superpartners by increasing the soft supersymmetry-breaking parameters. This is sensible, since in this limit the dimensionless, finite linear Yukawa coupling and the corresponding fermion mass $m_f$ must only depend on the electro-weak symmetry breaking scale.
\par


 \subsection{Cubic Yukawa Coupling } 
\par
Next, let us consider the cubic Yukawa term at dimension six 
$$
 - \eta_f ~H^\dagger H ~ \bar{f} H f ~+~ {\rm h.c.} 
$$
The topologies of Fig.\,\ref{fig:CubicYukop}, with three Higgs insertions is the diagram to be computed. This topology will give the cubic Yukawa coupling $\eta_f$. From the low-energy Higgs theorem, the contributions to the cubic Yukawa may be calculated directly from the linear Yukawa coupling and radiative fermion-mass expression of Eq.\,(\ref{massCoupling}), by repeatedly taking derivatives with respect to the Higgs vacuum expectation value.
\par 
Starting from the linear, mass-Yukawa loop-function 
$$
f_\lambda(x,y)= \frac{g(x)-g(y)}{x-y}
$$
where $x=m_1^2/m_\lambda^2$, $y=m_2^2/m_\lambda^2$ and $g(t)=\frac{2 t \log t}{(t-1)}$, we  apply the Higgs low-energy theorem  
\beq
2 \frac{\partial^2 f_\lambda(x,y)}{\partial v^2}\Big |_{v=0} = \left[ 4 \frac{\partial f_\lambda(x,y)}{\partial v^2}+ 4 v \frac{\partial}{\partial v}\left(\frac{\partial f_\lambda(x,y)}{\partial v^2}\right) \right] \Big |_{v=0} =4 \frac{\partial f_\lambda(x,y)}{\partial v^2} \Big |_{v=0} 
\eq
We remind ourselves that in our convention the Higgs field was defined as $H = {1 \over \sqrt{2} } \left( v + h \right) $, which gave the additional factor of 2 in the left hand side above.
\par
Now, we may apply the chain rule
\beq
\frac{\partial f_\lambda}{\partial v^2}(m_1^2/m_\lambda^2, m_2^2/m_\lambda^2)=\frac{1}{m_\lambda^2}\left[ \frac{\partial m_1^2}{\partial v^2}\frac{\partial f_\lambda(x,y)}{\partial x} +  \frac{\partial m_2^2}{\partial v^2}\frac{\partial f_\lambda(x,y)}{\partial y} \right] 
\eq
where
\beq
\frac{\partial f_\lambda(x,y)}{\partial x}=\frac{\partial_xg(x)}{(x-y)}-\frac{(g(x)-g(y))}{(x-y)^2} 
\eq
\beq
\frac{\partial f_\lambda(x,y)}{\partial y}=-\frac{\partial_yg(y)}{(x-y)}+\frac{(g(x)-g(y))}{(x-y)^2} 
\eq
\par
Also, as defined earlier, in our notations 
$$
m_1^2=\frac{1}{2}\left[2 m_L^2+\Delta_L+\Delta_R - \sqrt{(\Delta_L-\Delta_R)^2+ 4 \tilde{X}_f^2 }\right] \equiv \chi_1-\chi_2 
$$
$$
m_2^2=\frac{1}{2}\left[2 m_L^2+\Delta_L+\Delta_R + \sqrt{(\Delta_L-\Delta_R)^2+ 4 \tilde{X}_f^2 }\right] \equiv \chi_1+\chi_2 
$$
with the D-term factors given as before by, $\Delta_L=(T^f_{3L}-Q^f s_w^2)\,m_Z^2 \cos 2\beta$ and $\Delta_R=Q^f s_w^2\,m_Z^2 \cos 2\beta$; $T^f_{3L}$ is the third component of the weak-isospin and $Q_f$ is the electromagnetic charge. Also, as defined earlier, $\tilde{X}_f \equiv \tilde{A}_f \langle H^0 \rangle$. Note also that here we have already set $m_L=m_R$, to simplify subsequent expressions.
\par
Applying the derivatives give
\beq
\frac{\partial f_\lambda(x,y)}{\partial v^2} \Big |_{v=0}=\frac{1}{m_\lambda^2}\left[\frac{\partial \chi_1}{\partial v^2}  \left( {{\partial f_\lambda} \over {\partial x}} + {{\partial f_\lambda} \over {\partial y}}  \right) +    \frac{\partial \chi_2}{\partial v^2}  \left( {{\partial f_\lambda} \over {\partial y}} - {{\partial f_\lambda} \over {\partial x}}  \right) \right] \Big |_{v=0} 
\eq
\par
Using the definitions of $g(t)$ and the D-terms, we get for the first and second terms inside the square brackets,
\bea
\frac{\partial \chi_1}{\partial v^2}  \left( {{\partial f_\lambda} \over {\partial x}} + {{\partial f_\lambda} \over {\partial y}}  \right) \Big |_{v=0} &=& \left(\frac{\Delta_L + \Delta_R}{ v^2}\right)~\frac{(1-x^2+2 x \log x)}{x (x-1)^3}  \nn \\
\frac{\partial \chi_2}{\partial v^2}  \left( {{\partial f_\lambda} \over {\partial y}} - {{\partial f_\lambda} \over {\partial x}}  \right)\Big |_{v=0} &=& \left(\frac{\tilde{A}_f^2}{6 m_\lambda^2}\right)~\frac{1-6 x+3 x^2 +2 x^3-6 x^2 \log x}{x^2 (x-1)^4} 
\eea
\par
Putting everything together we have for $m_L=m_R$,
\beq
4 \frac{\partial f_\lambda(x,y)}{\partial v^2} \Big |_{v=0}=  {4 \over m_\lambda^2 }\left(\frac{\Delta_L + \Delta_R}{ v^2}\right)~\frac{(1-x^2+2 x \log x)}{x (x-1)^3}+{{2\tilde{A}_f^2} \over {3 m_\lambda^4} }~~\frac{1-6 x+3 x^2 +2 x^3-6 x^2 \log x}{x^2 (x-1)^4} 
\eq
Note that the D-term sum 
\beq
\Delta_L+\Delta_R=\pm {1 \over 2} m_Z^2 \cos 2 \beta= \pm {1 \over 8} (g^2+g^{\prime 2}) v^2 \cos 2\beta 
\eq
where $+$ is for the up-quarks and $-$ is for the down-quarks and charged-leptons. Though the $\Delta_L+\Delta_R$ factor gets contributions from both $SU(2)_L$ and $U(1)_Y$ D-terms, only the term proportional to the weak-isospin survives.
\par
Using the above results give, after normalizing the loop-functions, the cubic Yukawa coupling
\bea
\eta_f&=&\mp\frac{g^2 C}{6 (4\pi)^2}\frac{\tilde{A}_f}{m_\lambda m_R^2}~ (g^2+g^{\prime\, 2}) \cos 2\beta ~ f_\eta(m_L^2 / m^2_{\lambda} , m_R^2 / m^2_{\lambda})\nn \\
 &+& \frac{g^2 C}{3 (4\pi)^2}~\frac{\tilde{A}_f^3}{m_\lambda m_R^4}~ ~\tilde{f}_\eta(m_L^2 / m^2_{\lambda} , m_R^2 / m^2_{\lambda})  
 \label{CubicYukawaCoupling}
\eea

where
\bea
f_\eta(x,x) &=&  {3 \left( 1 - x^2 + 2 x \ln x \right) \over (1-x)^3 }  \nn \\
\tilde{f}_\eta(x,x) &=& \frac{2 (1-6 x +3 x^2+2 x^3-6 x^2 \log x)}{(x-1)^4} 
\eea
with $f_\eta(1,1)=1$ and $\tilde{f}_\eta(1,1)=1$. Observe that the term proportional to $\tilde{A}_f^3$ above, for $\tan\beta \geq 1$, adds destructively with the D-term dependent contributions for the down-quarks and charged leptons and constructively for the up-quarks; in Eq.\,(\ref{CubicYukawaCoupling}), after the function normalizations, the $-$ is now for the up-quarks and $+$ is for the down-quarks and charged-leptons. 
\par
To put this in a general context, and to understand why the three Higgs topology uniquely leads to $\eta_f$, note that for dimension-six operators, vector fermion currents form isospin singlets or triplets thereby only combining with even number of Higgs fields while scalar/tensor fermion currents form isospin doublets and may be paired with an odd number of Higgs fields. With this fact, combining three Higgs fields is unique due to one of the two doublets in this combination vanishing, by $\varphi^\dag \tilde{\varphi}=\epsilon_{\alpha \beta} \varphi^{*\,\alpha}\varphi^{*\,\beta}=0$~\cite{Grzadkowski:2010es}.
\par
The cubic Yukawa coupling along with the linear Yukawa coupling contributes to the effective mass Yukawa coupling, $ \lambda_f^{\text{\tiny{ eff}}}$, defined in Eq.\,(\ref{effmyukeffhyuk})
\ba
 \lambda_f^{\text{\tiny{ eff}}} &=&{ {g^2 C}\over 16 \pi^2}~{ \tilde{A}_f \over m_\lambda  } f_{\lambda}(m_{L}^2 / m^2_{\lambda} , m_{R}^2 / m^2_{\lambda})  \mp
{{g^2 C} \over 48 \pi^2}~{ \tilde{A}_f \over m_\lambda m^2_{f_R} } 
 m_Z^2 \cos 2\beta ~ f_\eta(m_{L}^2 / m^2_{\lambda} , m_{R}^2 / m^2_{\lambda})~~~\nn  \\ 
 &+& \frac{g^2 C}{96 \pi^2}~\frac{v^2 \tilde{A}_f^3}{m_\lambda m_R^4}~ ~\tilde{f}_\eta(m_L^2 / m^2_{\lambda} , m_R^2 / m^2_{\lambda}) 
 \ea
Any tree-level Yukawa coupling that may be present, is assumed to be small or vanishing. Note again that due to the different combinatorics, the cubic Yukawa term contributes with a different numerical coefficient to the mass-Yukawa coupling compared to the Higgs-Yukawa coupling. Therefore, already at this level, the expectation from SM for the equality of the mass-Yukawa and Higgs-Yukawa is no longer true.

\subsection{Yukawa Radius Coupling} 

\par
Finally, let us consider the Yukawa radius operator 
$$
- \zeta_f ~\bar{f} D^2 H  f ~+~ {\rm h.c.}  
$$
where $D$ is the covariant derivative,
$
D^2 H = \left( \Box + 
  2 i A_\mu \partial^\mu - 
   A_\mu A^\mu  
  \right) H 
$ with $ A_\mu = {1 \over 2} g' B_\mu + g W_\mu$ and
$\Box = \partial_\mu \partial^\mu$. 
\par
This is readily calculated from the Higgs-penguin three-point function, of Fig.\,\ref{fig:LinearYukop}, with non-zero Higgs external-momenta, by repeatedly taking derivatives with respect to the Higgs external-momenta. For reasons of brevity we only present the final result of the computation. To leading order, the result for the Yukawa radius coupling comes out to be 
\beq
\zeta_f =  
 {1 \over 24 \pi^2} ~ g^2 C ~ { \tilde{A}_f \over m_\lambda m_{R}^2}  ~  
~ f_{\zeta}(m_{L}^2 / m^2_{\lambda} , m_{R}^2 / m^2_{\lambda})  
\label{YukawaRadiusCoupling}
\eq
where 
\beq
f_\zeta(x,x) = { (1-x)(5 + (20-x)x) + 6x(3+x)\log x
 \over (1-x)^4 }  
\eq
and $f(1,1)=1$. Again, note that we have implicitly set $m_L=m_R$ to simplify the final expression.
\par
Note that the radiative fermion mass and the Higgs-Yukawa radius above, arise from the same topology of Fig.\,\ref{fig:LinearYukop}. Therefore the Higgs Yukawa-radius is effectively not suppressed by any additional loop-factor. This is distinct from the conventional tree-level Yukawa case, where the Yukawa radius is smaller by a loop-factor relative to the Compton wavelength of the contributing particles.
\par
The presence of the Higgs-Yukawa radius coupling, $\zeta_f$, implies that the difference in the mass-Yukawa coupling, $ \lambda_f^{\text{\tiny{ eff}}}$, and the Higgs-Yukawa coupling, $\lambda^h_f$, also contains a momentum dependent piece, apart from the combinatorial one due to the cubic Yukawa coupling, $\eta_f$, already discussed. Note that the Yukawa radius gives no contribution to the fermion masses, which are computed for $q\rightarrow 0$, when the external Higgs fields are replaced by their vacuum expectation values. Thus, interestingly, the Higgs-Yukawa couplings in radiative fermion mass frameworks are found to have a leading order $q^2$ dependence; this is over and in addition to the usual renormalization group evolution effects. Note that the function $f_\zeta(x,y) > 0$, and the contribution, for $\tan\beta \geq 1$, adds constructively with the D-term dependent cubic Yukawa term for the charged-lepton/down-quarks and destructively for the up-quarks. The Yukawa radius contribution always adds destructively with the $A_f^3$ dependent Cubic Yukawa term.

\section {Higgs Boson Yukawa Form Factors}
\label{sec:YukawaFormFactor}

\par
With the results of the previous section, we now have the ingredients required to compute the deviation of Higgs-Fermion couplings and anomalous magnetic moments, in the soft-Yukawa framework. 
\par
As motivated in \cite{CraigThomasThalapillil}, after the current run of the LHC, there is still room for sizable deviations in the couplings of Higgs to fermions. The deviations may be parametrized by the quantity 
\beq
\kappa_f \equiv { \lambda^h_f \over  \lambda_f^{\text{\tiny{ eff}}}}  
\eq
which in the SM would be unity. From a global fit to the available CMS and ATLAS Run I data, we found for instance that deviations of even $\sim \,50\%$ are currently allowed for the tau-lepton and bottom-quark couplings to the Higgs boson~ \cite{CraigThomasThalapillil}. The fits were performed with the large couplings to the top-quark and massive vector-bosons fixed, while allowing the small effective Higgs boson couplings to bottom quark, tau lepton, photon and gluon bi-linears to float.
\par
With substantial improvements expected at the next run of the LHC in measuring these couplings, it is projected that they could be measured to within an uncertainty of $\sim 1-5\%$~\cite{Peskin:2013xra}. With the planned ILC, colliding  $e^+e^-$ and running at $1\,\rm{TeV}$ with $1\,\rm{ab}^{-1}$ integrated luminosity, it is even speculated that the Higgs-fermion couplings could be measured to much less than $\sim 1\%$ uncertainties~\cite{Peskin:2013xra}. Given all these promising future projections it is prudent to investigate further the deviations one could potentially obtain in soft-Yukawa models.
\begin{figure}[p!]
\begin{center}
\includegraphics[width=12.0cm]{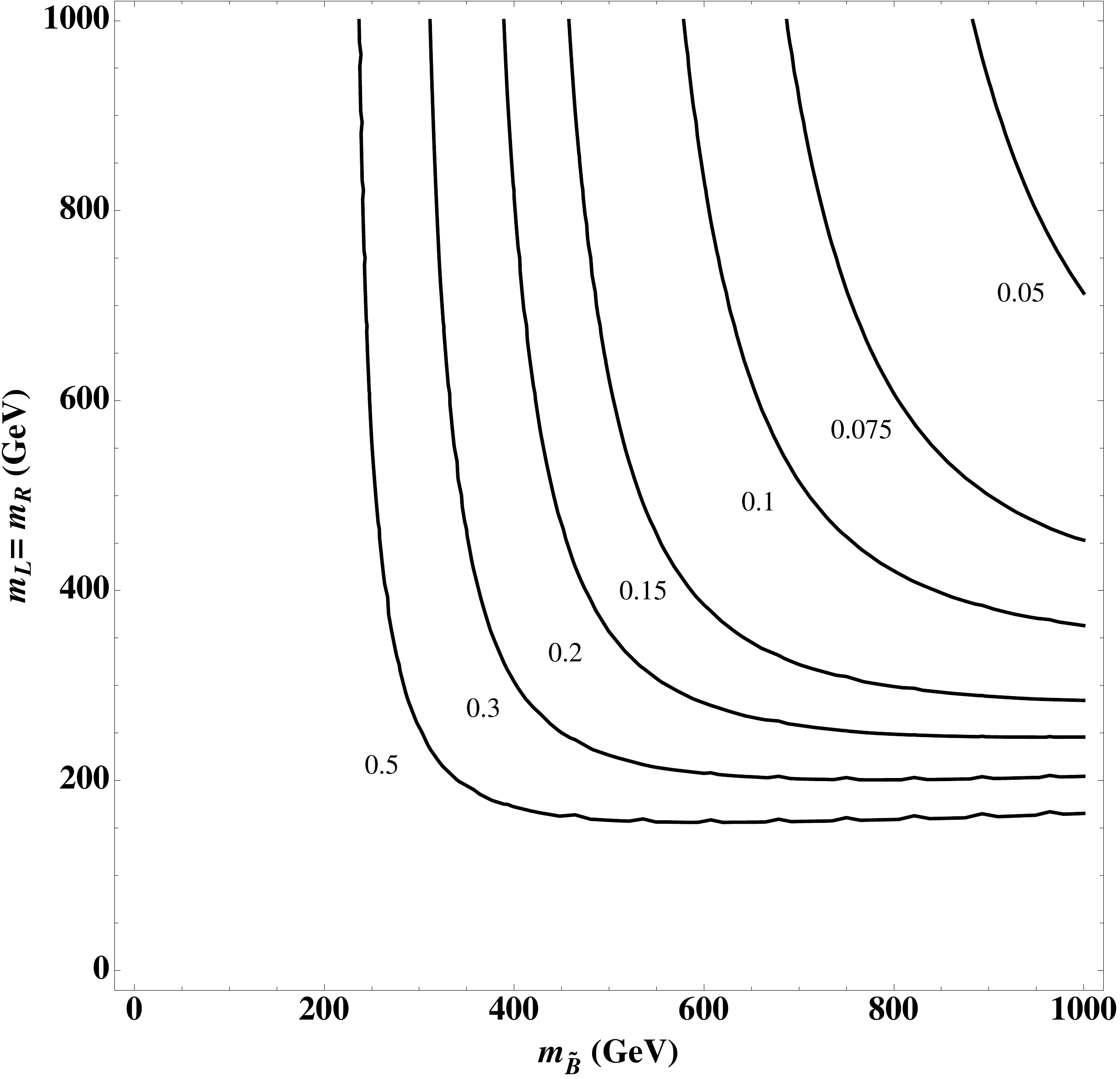} 
\caption{ Deviations in Higgs-Muon couplings, $\kappa_{\mu} -1$, relative to SM expectations, in soft-Yukawa models of radiative fermion mass generation. The plot shown is for $\tan\beta=1$ and the gaugino is assumed to be the lightest-bino. Notice that even for modest values of sparticle masses, around a TeV, the deviations in the Higgs-muon couplings due to the Higgs-Yukawa form factor could be of the order of a few percent.}
\label{fig:DeltaKappamuZoomed}
\end{center}
\end{figure}

\begin{figure}[p!]
\begin{center}
\includegraphics[width=12.0cm]{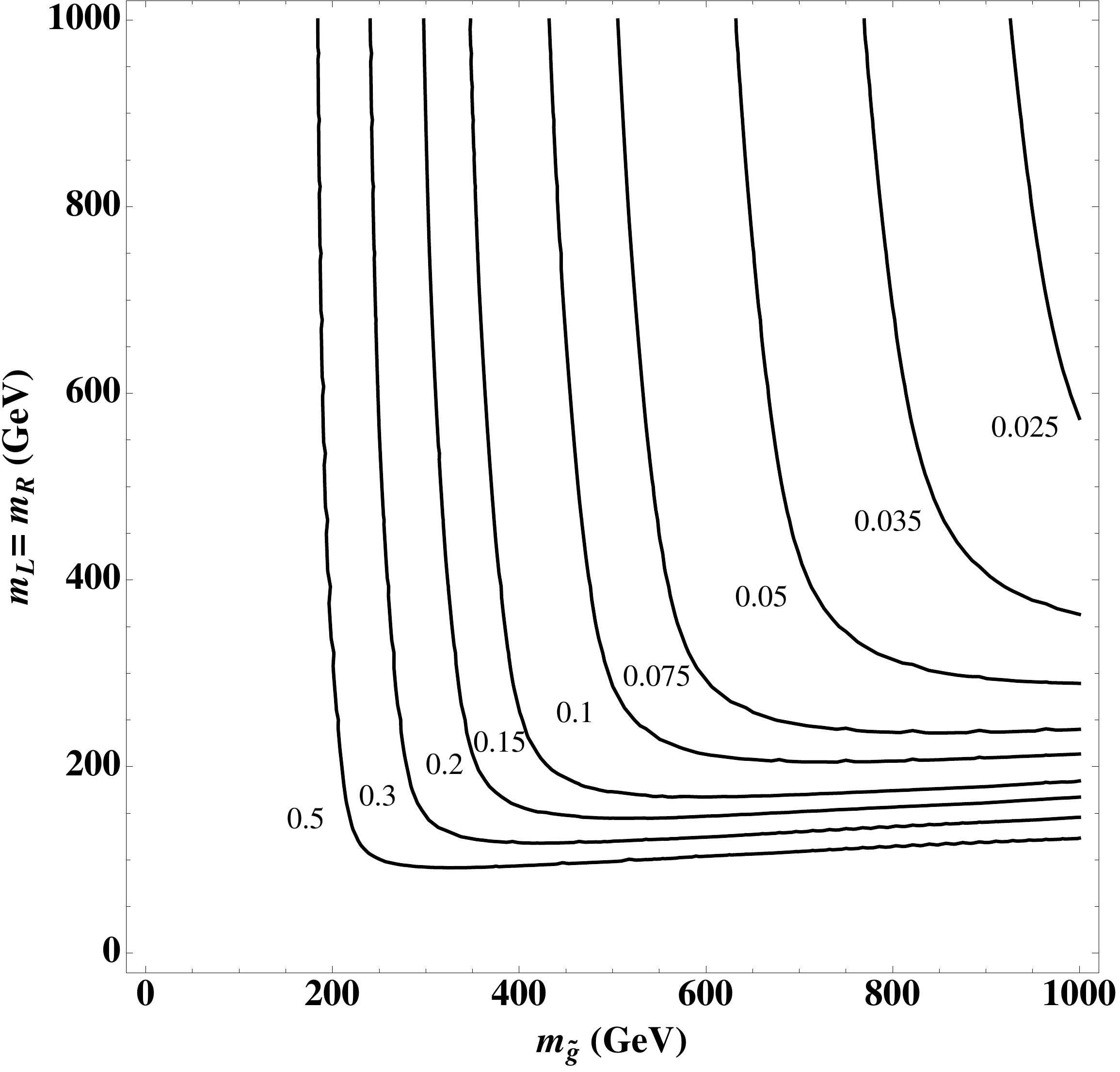} 
\caption{ Deviations in Higgs-bottom quark couplings parametrized by $\kappa_{b} -1$ in soft-Yukawa models. Again, the plot is shown for $\tan\beta=1$ with the gaugino being the gluino. Once again, observe that even for sparticle masses, around a TeV, the deviations in the Higgs-bottom couplings due to the Higgs-Yukawa form factor are relatively sizable.}
\label{fig:DeltaKappabZoomed}
\end{center}
\end{figure}

\par
From Eqs.\,(\ref{CubicYukawaCoupling}) and (\ref{YukawaRadiusCoupling}), with the basis independent combination from Eq.\,(\ref{kappaops}), the Higgs-Yukawa coupling is given by 

\bea
\lambda^h_f &=&~{{g^2 C} \over 16 \pi^2} { \tilde{A}_f \over m_\lambda  } f_{\lambda}(m_{L}^2 / m^2_{\lambda} , m_{R}^2 / m^2_{\lambda})\mp~ {{g^2 C } \over 16 \pi^2} ~ { {\widetilde{A}_f} \over m_\lambda m_{R}^2} ~  m_Z^2 \cos 2 \beta 
~ f_{\eta}(m_{L}^2 / m^2_{\lambda} , m_{R}^2 / m^2_{\lambda})~ \nn \\ 
 &+& \frac{g^2 C}{32 \pi^2}~\frac{v^2 \tilde{A}_f^3}{m_\lambda m_R^4}~ ~\tilde{f}_\eta(m_L^2 / m^2_{\lambda} , m_R^2 / m^2_{\lambda})-~{{ g^2 C} \over 24 \pi^2}~ { {\tilde{A}_f m_h^2} \over m_\lambda m_{R}^2}~f_{\zeta}(m_{L}^2 / m^2_{\lambda} , m_{R}^2 / m^2_{\lambda})  
\label{HiggsYukawacpl}
\eea
The mass-Yukawa coupling is similarly given by
\bea
  \lambda_f^{\text{\tiny{ eff}}} &=&~{ {g^2 C}\over 16 \pi^2}~{ \tilde{A}_f \over m_\lambda  } f_{\lambda}(m_{L}^2 / m^2_{\lambda} , m_{R}^2 / m^2_{\lambda})  \mp
{{g^2 C} \over 48 \pi^2}~{ \tilde{A}_f \over m_\lambda m^2_{f_R} } 
 m_Z^2 \cos 2\beta ~ f_\eta(m_{L}^2 / m^2_{\lambda} , m_{R}^2 / m^2_{\lambda})~~~\nn  \\ 
 &+& \frac{g^2 C}{96 \pi^2}~\frac{v^2 \tilde{A}_f^3}{m_\lambda m_R^4}~ ~\tilde{f}_\eta(m_L^2 / m^2_{\lambda} , m_R^2 / m^2_{\lambda})  
  \label{MassYukawacpl}
 \eea
 Again, any tree-level Yukawa coupling is assumed to be relatively small or vanishing. If the tree-level Yukawa is vanishing, then from the above expressions and Eq.\,(\ref{kappaops}) we get 
\beq
 \kappa_f = 1 - {2 m_h^2 \over 3 m_{R}^2} 
\left[  
 { f_\zeta (x,y) 
  \pm  \cos 2 \beta (m_Z^2 / m_h^2) f_\eta(x,y)
  -{1 \over 2} (v^2/m_h^2) (\tilde{A}^2/m_R^2) \tilde{f}_\eta(x,y)
  \over 
  f_\lambda(x,y)  } 
   \right]  
\label{kappafermion}
\eq
where $x=m_L^2 / m^2_{\lambda} $, $y=m_R^2 / m^2_{\lambda} $, and we have set $m_L=m_R=\tilde{m}$ implicitly.
\par
The quantity $\kappa_f -1$ parametrizes the deviations from the SM Higgs-fermion couplings. As the sfermion spectrum gets very heavy note that $\kappa_f\rightarrow 1 $ as one should expect. In this limit, the effective mass-Yukawa and Higgs-Yukawa couplings coincide.
\par
Note that the final expression for $\kappa_f$ is independent of the gauge-factor $C$. The first term inside the brackets of Eq.\,(\ref{kappafermion}), coming from the Yukawa radius, is always positive. The second term from the D-term dependent cubic Yukawa contribution, for $\tan\beta \geq 1$ , has an overall negative sign for the up-quarks and an overall positive sign for the down-quarks/charged-leptons. The final term is the one from the $A_f^3$ dependent cubic Yukawa coupling, is negative and destructively adds to the Yukawa radius. 
\par
We wish to understand the interplay of the various contributions and how they interfere with each other. Towards this end, we will have to solve the fermion mass equation, Eq.\,(\ref{massCoupling}), for $\tilde{A}_f$, at each point in the $(m_\lambda,\,\tilde{m})$ parameter space. This will impose the constraint that one gets a viable value for the fermion mass at that point. Note that the above equation is a non-linear function of $\tilde{A}_f$, since the loop function contains scalar mass eigenstates that implicitly contain a resummation of an arbitrary number of chiral insertions. To leading order though, this may be approximated to a cubic equation which gives the correct solution. We solve this equation at every point in the parameter space and pick the real-valued solution for $\tilde{A}_f$. This is then used to compute the value of $\kappa_f$ from Eq.\,(\ref{kappafermion}).
\par
With the above procedure we compute the deviation, $\Delta \kappa_f=\kappa_f-1$, for the interesting case of the muon and the bottom-quark. The deviations in the Higgs-boson and fermion couplings, for various values of the scalar and gaugino mass-parameters, are shown in Figs.\,\ref{fig:DeltaKappamuZoomed} and \ref{fig:DeltaKappabZoomed} for the muon and the bottom-quark. In our case- with soft, radiatively generated fermion masses- we observe that the overall effect of the dimension-six operators is to enhance these Higgs-fermion couplings relative to the standard model, $\kappa_{\mu/b}~\geq~1 $. Even for conservative values of the sparticle masses, one observes that, relatively large deviations are possible in the couplings. In scenarios where a lighter sparticle spectrum is still allowed, one even encounters deviations that are potentially $\mathcal{O}(10\%)$ or larger. As an aside, also note that the modified Higgs coupling to fermion bi-linears still have the same kinematic structure as the renormalizable SM couplings, and therefore will modify just the rate measurements. Thus, these effects could readily show up in LHC Higgs measurements in the near future. Finally we comment that for large $A_f$ solutions, in the context of vacuum stability, we have already seen from Figs.\, \ref{fig:AbyMMuon} and \ref{fig:AbyMBottom}, that viable swaths of the parameter space are present that are metastable. In these regions one obtains realistic muon/bottom-quark masses while also having potentially large corrections to the Higgs-fermion couplings through the Higgs-Yukawa form factors.

\section {Anomalous Magnetic Moments in Soft-Yukawa Models}
\label{sec:AnomalousMagneticMoment} 

\begin{figure}[htb]
\begin{center}
\includegraphics[width=8.0cm]{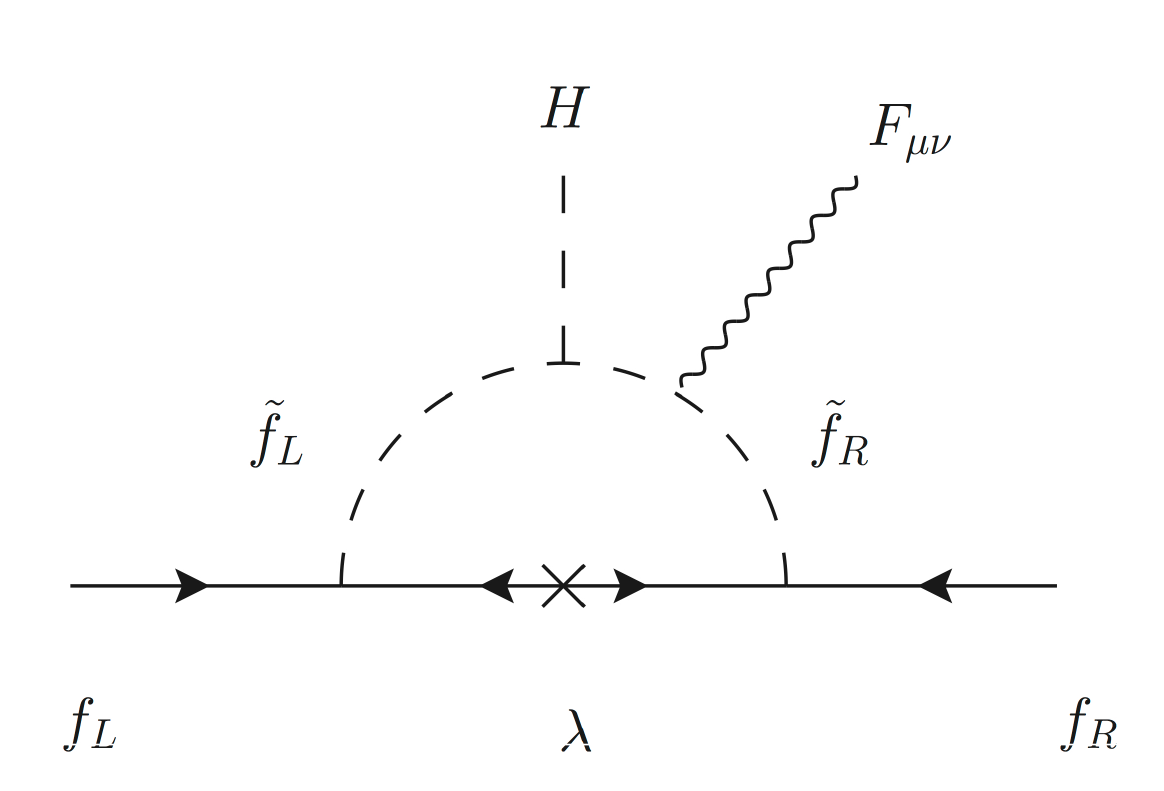} 
\caption{One-loop supersymmetric contribution to the dimension-six, fermion electromagnetic dipole operator. With a Higgs expectation value, note that this diagram will contribute to fermion anomalous magnetic moments in models with soft-Yukawa couplings. Since the fermion mass in soft-Yukawa models are generated at one loop, the above contribution to the fermion electromagnetic dipole operator is parametrically of the same order.}
\label{fig:HiggsDipoleops}
\end{center}
\end{figure}

We will investigate the contributions to anomalous magnetic moments in this section. We will focus specifically on the interesting case of the muon anomalous magnetic moment. There currently exists a $\sim\,3\sigma$ discrepancy here between experiment and theory. It will therefore be interesting to consider the contribution from the soft-Yukawa models to $a_\mu$. We will comment on the status of other lepton and quark magnetic moments at the end.
\par
The anomalous magnetic moment of the muon has been measured to very good precision. Current measurements yield a value- $\left[a_{\mu}\right]_{ \text{\tiny exp}}=(11\,659\,2091 \pm 54_{\rm stat} \pm 33_{\rm syst})\times 10^{-11}$~\cite{{Bennett:2006fi},{Mohr:2012tt}}. Comparing the theoretical predictions and the current experimental result gives a discrepancy between experiment and theory of~(Please see~\cite{Agashe:2014kda} and relevant references therein)
\begin{equation}
\begin{array}{l}
\left[\Delta a_{\mu}\right]^{\text{\tiny exp.}}_{\text{\tiny th}.}=~(288\pm 80)\times 10^{-11}  
\end{array}
\end{equation}

\par
This is a relatively sizable discrepancy, $\sim\,3\sigma$, which has persisted over time. In the context of supersymmetric theories it has been pointed out that such a discrepancy could be the result of additional contributions that may be $\tan\beta$ enhanced~\cite{{Moroi:1995yh},{Martin:2001st},{Graesser:2001ec}} . We wish to investigate what the generic predictions are for $a_\mu$ in the context of radiatively generated soft-Yukawa terms.

\par
Towards this end, consider now the dimension-six, fermion electromagnetic dipole Lagrangian operator 
\beq
-~{ \xi_f~\bar{f} \sigma^{\mu \nu} H f ~F_{\mu \nu}} ~+~ {\rm h.c.} 
\eq
where $f,\,\bar{f}$ are two-component fermions in the Weyl basis. This may be generated from a topology such as the one shown in Fig.\,\ref{fig:HiggsDipoleops}. This leads to a magnetic-dipole moment (MDM) term $ - \xi_f^\prime~\bar{\Psi} \sigma^{\mu \nu} H \Psi~F_{\mu \nu}$ where,  $\Psi,\,\bar{\Psi}$ are four-component fermions and $\sigma^{\mu \nu}  = {i \over 2} [ \gamma^\mu , \gamma^\nu ]$. For brevity, we have defined $\xi_f^\prime=\,\Re(\xi_f)$ which is related to the anomalous magnetic moment through $\xi_f^\prime \frac{v}{\sqrt{2}}=\frac{e Q_f}{4 m_f} a_f $, implying that 
\beq
a_f =\frac{2\sqrt{2} v m_f}{e Q_f} \xi_f^\prime 
\label{xirel}
\eq
Note also that in our conventions, $Q_\mu=-1$ and $e$ is the magnitude of the electric charge.
\par
From the photon-penguin topology with an additional Higgs insertion, of Fig.\,\ref{fig:HiggsDipoleops}, the quantity $\xi_f^\prime$ may be calculated. Specializing to the case of the muon, it has the form
\beq
\xi_\mu^\prime =   -{\alpha' \over 96 \pi}\, e Q_\mu \,{ \tilde{A}_\mu \over 
   m^3_\lambda } 
~  f_\xi(m_1^2 /m_\lambda^2, m_2^2 / m_\lambda^2) 
\eq

 where 
\beq
f_\xi(x,y) = \frac{ g_\xi(x)-g_\xi(y)}{(y- x)} 
\eq
with
\beq
g_\xi(x)=\frac{6}{(x-1)^3}\left( -1+x^2-2 x \log x \right) 
\eq
and 
\beq
f_\xi(x,x)=\frac{6}{(x-1)^4}(-5+4 x + x^2-(4x+2)\log x) 
\eq
where $f_\xi(1,1)=1$. 
\par
Using the expression for $m_\mu$, from Eq.\,(\ref{muonmassCoupling}), and Eq. (\ref{xirel}), we can now rewrite $\xi_\mu^\prime$ as an anomalous magnetic moment in soft-Yukawa models
\beq
a^{\text{\tiny{SUSY}}}_\mu = + {m_\mu^2  \over 3 m_{\tilde{B}}^2 } ~
{  f_\xi(m_{1}^2 / m_{\tilde{B}}^2 , m_{2}^2 / m_{\tilde{B}}^2 ) \over 
f_\lambda(m_{1}^2 / m_{\tilde{B}}^2 , m_{2}^2 / m_{\tilde{B}}^2 )  } 
\eq
$\tilde{B}$ is considered to be the lightest-bino contributing most significantly to the process.
\par
Note that since $f_\xi(x,y) > 0$ and $f_\lambda(x,y) > 0$ everywhere in the physical region of interest, this is a positive contribution to the anomalous magnetic moment. This is therefore in the same direction as the current $\sim 3\sigma$ discrepancy in $a_\mu$ and is a prediction of the soft-Yukawa model. This was also previously noted in~\cite{Borzumati:1999sp}.
\begin{figure}[p!]
\begin{center}
\includegraphics[width=13.0cm]{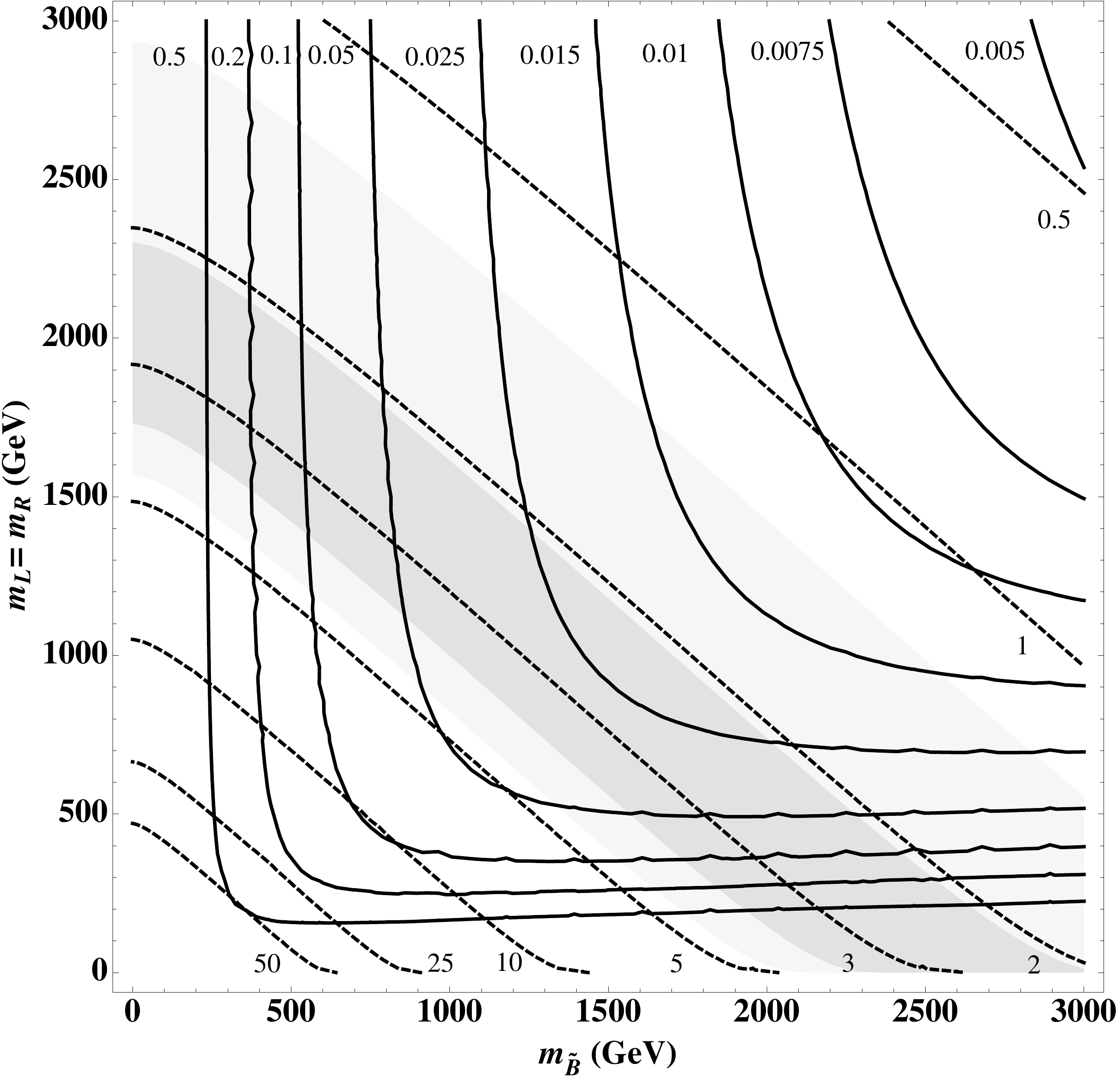} 
\caption{Contributions to $a_\mu$ in the Soft-Yukawa model (dashed lines) are shown overlaid on values of $\Delta\kappa_\mu$, the muon Higgs-Yukawa form factor (solid lines). The numbers labelling the dashed contours correspond to $(a^{\text{\tiny{SUSY}}}_\mu \times10^{9})$. The $1\sigma$ and $2\sigma$ contours, signifying the current discrepancy between experiment and theory, in $a_\mu$ are illustrated by the dark-grey and light-grey bands respectively. The gaugino is assumed to be the lightest-bino. Note that the contributions to $a_\mu$ are strictly positive and therefore has the correct sign to potentially explain the current discrepancy, between theory and experiment. The $\Delta \kappa_\mu$ contours were made assuming $\tan\beta=1$. We see that in the regions where the current discrepancy in $a_\mu$ could be expalined by the soft-Yukawa model, the deviations due to the Higgs-Yukawa form factors could also be sizable enough to be probed in the near future.}
\label{fig:DeltaMuonMagMoment}
\end{center}
\end{figure}

\par
In Fig.\,\ref{fig:DeltaMuonMagMoment}, we plot $a^{\text{\tiny{SUSY}}}_\mu$ contours overlaid over contours of the muon Higgs-Yukawa formfactor $\Delta \kappa_\mu$. The plots were made as a function of $m_{\tilde{B}}$ (taken to be the lightest bino) and assuming $m_L=m_R$, as before. In the most interesting regions, consistent with an explanation for the current $a_\mu$ discrepancy at the $95\%\,\rm{C.L.}$, we note from Figs.\,\ref{fig:DeltaMuonMagMoment} that the corresponding deviation in the Higgs-muon coupling could be significant. With a more liberal $3\sigma$ band one could expect even larger $\mathcal{O}(10\%)$ or higher deviations.  The $\Delta \kappa_\mu$ contours were made assuming $\tan\beta=1$, and solving again the radiative muon mass equation at each point in parameter space, to get a viable value for the trilinear term, as in Fig.\,\ref{fig:AbyMMuon}.
\par
It is interesting to note that the contribution to $a_\mu$ could be $\mathcal{O}(1)$ and large in these models. In the context of a radiatively generated muon mass, we do not pay any extra price in terms of loop-factors for $a^{\text{\tiny{SUSY}}}_\mu$. Both the muon mass and the muon anomalous magnetic moment are generated at the same loop level. It is also to be noted that in the soft-Yukawa framework, unlike the usual case in the MSSM, $a_\mu$ is independent of $\tan\beta$ at leading order. There is no $\tan \beta$ enhancement per se, but the contribution may still be significant due to the fact that the fermion mass and anomalous magnetic moment arise at the same loop-order. The presence or absence of a discrepancy in $a_\mu$, may be further probed in upcoming experiments in the near future~\cite{Venanzoni:2014ixa}. This makes soft-Yukawa models particularly interesting.
\par
 It must also be commented that due to the contributions to $a_\mu$ being proportional to $m_l^2$, the corresponding value for the electron anomalous magnetic moment will be heavily suppressed. Currently, the most accurate measurement of $a_e$ gives  $\left[ a_e\right]_{ \text{\tiny exp}} = (115965218.07\pm 0.03 )\times 10^{-11}$~\cite{{2008PhRvL.100l0801H},{PRA.Hanneke} }. The current SM theoretical prediction is $ \left[ a_e\right]_{ \text{\tiny th}} = (115965218.01\pm0.08)\times 10^{-11}$ (see for example the relevant chapter in \cite{quintvogel}) . In the model, all things being equal, the corresponding contribution to $a_e$ will be suppressed at least by $m_e^2/m_\mu^2\sim 10^{-5}$ and can be readily accommodated in the above uncertainties. Currently, the best limits on $a_\tau$ come from the DELPHI experiment at LEP2~\cite{Abdallah:2003xd}. They studied tau-pair production in the channel, $e^- ~e^+~\rightarrow ~e^-~e^+~\tau^-~\tau^+ $, at an integrated luminosity of $650~\rm{pb}^{-1}$. For the relevant energies at LEP2, $\sqrt{s}\in[183,\,208]\,\rm{GeV}$, the tau-production is dominated by a multiperipheral topology, relative to pair-production via Brehmmstrahlung. The $95\%$ C.L.  values obtained were $\left[ a_\tau\right]_{ \text{\tiny exp.\,range}}~\in~[-0.052,\,0.013] $, with a central fit value of $\left[a_\tau\right]_{ \text{\tiny exp}} = -0.018\pm0.017 $~\cite{Abdallah:2003xd}. The uncertainties in $a_\tau$ are clearly very large (larger than even the 1-loop prediction $\alpha_{\text{\tiny{EM}}}/2 \pi$ of QED) and thus it is hard to extract any meaningful constraints from $a_\tau$ at this point in time.
\par
In the quark sector, again potentially the most interesting case is that of the second and third generation quarks. Unfortunately, at this point in time the MDM constraints are relatively weak for the second and third generations. It was shown in~\cite{Kamenik:2011dk} that one can place limits on the top magnetic and chromo-magnetic dipole operator from its contribution to flavor changing processes (such as $b\rightarrow s \gamma$ for instance), and the measured $t\bar{t}$ cross-section. They found the  bounds $\mu_t^{\text{\tiny{MDM}}}  < 1.73\times10^{-3}\,\rm{GeV}^{-1}$ and $\mu_t^{\text{\tiny{CMDM}}}  < 2.9\times10^{-4}\,\rm{GeV}^{-1}$.
There is also a very recent result from the CMS collaboration placing limits on the top chromo-magnetic dipole moment, from the distribution of di-leptonic $t \bar{t}$ events at $7\,\rm{TeV}$~\cite{CMS:2014bea}. They place a limit $\Re[\mu_t^{\text{\tiny{CMDM}}}] ~\in~[-2.5\times 10^{-1}\,\rm{GeV}^{-1}, 6.8\times 10^{-4}\,\rm{GeV}^{-1}]$. We have assumed a top mass of $m_t = 172.4\,\rm{GeV}$, to convert the CMS limit into numbers consistent with the conventions of~\cite{Kamenik:2011dk}, for comparison. Measurements of the proton and neutron magnetic moments~\cite{Mohr:2012tt} suggests that the up-quark and down-quark magnetic moments are approximately $\mu_u^{\text{\tiny{MDM}}} \sim 0.3\,\rm{GeV}^{-1}$ and $\mu_d^{\text{\tiny{MDM}}} \sim -0.2\,\rm{GeV}^{-1}$. There are currently no definite measurements for the bottom-quark magnetic and chromo-magnetic dipole moments.


\section{Conclusions} 
\label{sec:conclusions}
\par
The recent discovery of the Higgs-like boson has opened up new experimental opportunities to probe for extensions of the SM. In this context it is therefore more pertinent to explore models where properties of the Higgs boson couplings or other Higgs-related observables may undergo modifications.
\par
An interesting possibility that we considered in this work, is that chiral violation in the supersymmetric sector could be, completely or partially, the origin of fermion masses. This is a plausible scenario that may be realized readily for the first and second generation quark, lepton masses and also for the bottom-quark mass. We briefly considered the validity of the framework in terms of vacuum stability. The relatively large trilinear terms required for the soft-Yukawa generation could potentially lead to color or charge breaking. We found that most of the first and second generation fermion masses can be accommodated in a consistent manner with a metastable vacuum. We focused in more detail on the more interesting and non-trivial cases of the muon and bottom-quark, and in these cases we identified slivers in the parameter space where the respective fermion masses could be generated solely  through radiative processes.
\par
A generic feature of the soft-Yukawa models we considered is that the couplings between the Higgs boson and fermions are modified from their SM values. These deviations, even for modest values of sparticle masses, could be sizable - sometimes by even $\mathcal{O}(1)$ factors in certain regions. With the projected uncertainties in the Higgs-fermion couplings expected to reduce, at the LHC and future $e^+e^-$ linear-colliders, to $\mathcal{O}(1-5\%)$ and $\mathcal{O}(0.1\%)$, these deviations could be probed in the very near future.
\par
The other interesting feature of these models is that they could give significant contributions to anomalous magnetic moments. We investigated the case of the anomalous magnetic-moment of the muon, which currently exhibits a $\sim3\,\sigma$ discrepancy between experiment and theory. We find that unlike the well-known MSSM case where $a_\mu$ is enhanced by $\tan\beta$ and potentially large, in the soft-Yukawa case the contribution is independent of $\tan\beta$ to leading order but still sizable. The large contribution primarily stems from the fact that the anomalous magnetic moment arises at the same loop-level as the fermion mass, unlike the case with tree-level Yukawas where it is suppressed by an additional loop. We discovered that the contribution is positive, in the same direction as the discrepancy, and that contributions well within $2\sigma$ of the current discrepancy may be easily accommodated. In these regions the Higgs Yukawa factors are also such that they are amenable to experimental discovery in the near future.

\vspace{0.5in}

{ \Large \bf Acknowledgments}

\smallskip \smallskip

\noindent
We would like to thank Nathaniel Craig for useful comments. 
This work was supported in part by the US Department 
of Energy under grant DOE-SC0010008.



\end{document}